\documentclass[12pt,a4paper]{article}
\usepackage{amssymb,amsfonts,amsmath}

\input{tcilatex}
\begin{document}

\title{Bell's nonlocality in a general nonsignaling case: quantitatively and
conceptually}
\author{Elena R. Loubenets \\
%EndAName
National Research University Higher School of Economics, \\
Moscow 101000, Russia}
\maketitle

\abstract{Quantum violation of Bell inequalities is now used in many quantum information applications and it is
important to analyze it both quantitatively and conceptually. In the present paper, we analyze violation of multipartite 
Bell inequalities via the local probability model --  the LqHV (local quasi hidden variable) model [Loubenets, 
J. Math. Phys. 53, 022201 (2012)], incorporating the LHV model only as a particular case and 
correctly reproducing the  probabilistic description of every quantum correlation scenario, more generally, every  
nonsignaling scenario. The LqHV probability framework allows us to construct nonsignaling analogs of 
Bell inequalities and to specify parameters quantifying violation of Bell inequalities -- Bell's nonlocality -- in a general 
nonsignaling case. For quantum correlation scenarios on an N-qudit state, we evaluate these nonlocality parameters 
analytically  in terms of dilation characteristics of an N-qudit state and also, numerically -- in d and N. In view of our 
rigorous mathematical description of Bell's nonlocality in a general nonsignaling case via the local probability model, 
we argue that violation of Bell inequalities in a quantum case is not due to violation of the Einstein-Podolsky-Rosen 
(EPR) locality conjectured by Bell but due to the improper HV modelling of "quantum realism".}%
\medskip

\noindent \textbf{Keywords: }Nonsignaling -- Bell's nonlocality -- The LqHV
modelling -- Quantum realism \bigskip

\section{Introduction}

In more than 50 years since the seminal paper \cite{1} of Bell, there is
still no a unique conceptual view\footnote{%
See Introductions in \cite{4, 5, 6, 7, 8} and discussions in \cite{9.1, 9.2,
9, 10, 11, 12}.}\ on quantum nonlocality conjectured by Bell \cite{2, 3} for
explaining quantum violation of the local hidden variable (LHV) statistical
constraints. However, in quantum information, nonlocality of a multipartite
quantum state is defined purely mathematically -- via violation by this
state of a Bell inequality and it is specifically in this sense quantum
nonlocality is now used in all experimental quantum information processing
tasks.

Moreover, from the practical point of view, it is also important to know
violation or nonviolation by an $N$-partite quantum state of Bell
inequalities of some specific class, hence, nonlocality or locality of an $N$%
-partite state under the corresponding class of correlation scenarios, for
example, under correlation scenarios with some specific numbers $%
S_{1},...,S_{N}$ of settings at $N$ sites. The latter type of \emph{partial}
locality of an $N$-partite quantum state, \emph{the} $S_{1}\times \cdots
\times S_{N}$\emph{-setting locality}, was analyzed in \cite{5, 6, 13, 14,
15}.

However, in all cases, \emph{quantifying nonlocality of an N-partite quantum
state, }full or partial, is associated with finding the maximal violation by
this state of the corresponding class of Bell inequalities. In the
literature, the well-known attainable upper bounds \cite{16, 17, 18, 19} on
quantum violation of specific Bell inequalities concern the
Clauser-Horne-Shimony-Holt (CHSH) inequality and the Mermin-Klyshko
inequality. It is also well known that the maximal quantum violation of
bipartite Bell inequalities on correlation functions cannot exceed the real
Grothendieck's constant $K_{G}^{(\mathbb{R})}\in \lbrack 1.676,1.783]$
independently of a dimension of a bipartite state and numbers of measurement
settings and outcomes per site. But the latter is not already the case for
quantum violation of bipartite Bell inequalities on joint probabilities and
last years bounds on the maximal quantum violation of \emph{general}%
\footnote{%
That is, Bell inequalities of an arbitrary type -- either on correlation
functions or on joint probabilities or of a more complicated form.} Bell
inequalities were intensively discussed in the literature, see \cite{6, 8,
19.2, 19.3, 19.4, 19.5, 19.6, 20} and references therein.

In the sense of violation of a Bell inequality, nonlocality is also inherent
to a general nonsignaling\footnote{%
On this notion, see section 3 in \cite{5}, also, section 2 below.}
correlation scenario and, in this case, we refer \cite{5} to it as \emph{%
Bell's nonlocality}. As we analyzed this mathematically in \cite{5}, for an
arbitrary correlation scenario, Bell's locality (in the sense of
nonviolation of all general Bell inequalities) implies the EPR
(Einstein--Podolsky--Rosen) locality \cite{30} and the EPR locality implies
nonsignaling -- but not vice versa. Therefore, \emph{Bell's nonlocality does
not necessarily lead to violation of the EPR locality.}

In the present paper, we analyze Bell's nonlocality via the local
probability model, \emph{the LqHV (local quasi hidden variable), }introduced
in\emph{\ }\cite{6, 21, 22} and incorporating the LHV probability model only
as a particular case. The LqHV model correctly reproduces the probabilistic
description of every nonsignaling correlation scenario (in particular, every
quantum scenario) and this allows us to construct nonsignaling analogs of
Bell inequalities and to specify parameters quantifying Bell's nonlocality,
partial and full, in a general nonsignaling case. For quantum correlation
scenarios on an $N$-partite state, we evaluate these nonlocality parameters
analytically via dilation characteristics of an $N$-partite state. For an $N$%
-qudit state, we also evaluate them numerically -- in $d,$ $S$ and $N$.

\emph{In view of our rigorous mathematical description of Bell's nonlocality
in a general nonsignaling case via the local probability model, we argue
that violation of Bell inequalities in a quantum case is not due to
violation of the EPR\ locality conjectured by Bell \cite{2, 3} but due to
the improper HV\ modelling of "quantum realism".}

The paper is organized as follows.

In Section 2, we introduce the functional approach \cite{23} to constructing
general multipartite Bell inequalities for arbitrary numbers of settings and
outcomes at each site and present the single general representation
incorporating in a unique manner all Bell inequalities. This representation
allows us to specify in section 4 violation of a Bell inequality in an
arbitrary nonsignaling case.

In Section 3, we introduce the notion of a LqHV (local quasi hidden
variable) probability model \cite{6, 21, 22} and discuss its validity for a
general correlation scenario.

In Section 4, we find nonsignaling analogs of Bell inequalities and specify
parameters quantifying Bell's nonlocality, partial and full, in a general
nonsignaling case.

In Section 5, for quantum correlation scenarios on an $N$-qudit state, we
evaluate these nonlocality parameters analytically and numerically. As an
example, we specify the quantum analog of the bipartite Bell inequality
presented in \cite{24}.

In Section 6, we discuss the conceptual issues of Bell's nonlocality.

\section{Multipartite Bell inequalities for arbitrary numbers of settings
and outcomes per site}

In this section, we present \emph{the functional approach }\cite{23} to
constructing \emph{general} multipartite Bell inequalities. In contrast to
the polytope approach \cite{16, 18, 25}, which is valuable for finding Bell
inequalities on correlation functions and joint probabilities in case of
small numbers of settings and outcomes per site, the functional approach
leads to a single general representation for Bell inequalities of any type
with arbitrary numbers of settings and outcomes at each site. This allows us
further easily to analyze in section 4 a modification of this general
representation for an arbitrary nonsignaling case.

Consider\footnote{%
For the general framework on the probabilistic description of multipartite
correlation scenarios, see \cite{5}.} an $N$-partite correlation scenario,
where each $n$-th of $N\geq 2$ parties (players) performs $S_{n}\geq 1$
measurements with outcomes $\lambda _{n}\in \Lambda _{n}$ of any nature and
an arbitrary spectral type. We label each measurement at $n$-th site by a
positive integer $s_{n}=1,...,S_{n}$ and each $N$-partite joint measurement,
induced by this correlation scenario and with outcomes 
\begin{equation}
(\lambda _{1},\ldots ,\lambda _{N})\in \Lambda =\Lambda _{1}\times \cdots
\times \Lambda _{N},  \label{0}
\end{equation}%
by an $N$-tuple $(s_{1},...,s_{N}),$ where $n$-th component specifies a
measurement at $n$-th site.

For concreteness, we further denote by $\mathcal{E}_{S,\Lambda },$ $%
S=S_{1}\times \cdots \times S_{N},$ an $S_{1}\times \cdots \times S_{N}$%
-setting correlation scenario with outcomes in $\Lambda $ and by $%
P_{(s_{1},...,s_{N})}^{(\mathcal{E}_{S,\Lambda })}$ -- a joint probability
distribution of outcomes under an $N$-partite joint measurement $%
(s_{1},...,s_{N})$ of a scenario $\mathcal{E}_{S,\Lambda }$.

The superscript $\mathcal{E}_{S,\Lambda }$ at notation $%
P_{(s_{1},...,s_{N})}^{(\mathcal{E}_{S,{\large \Lambda }})}$ indicates that,
under a scenario $\mathcal{E}_{S,\Lambda },$ this joint probability
distribution may depend not only on parties' settings $(s_{1},...,s_{N}),$
specifying this joint measurement, but also on settings of all (or some)
other measurements of this scenario. This is, for example, the case for
scenarios with two-sided memory \cite{26}.

A correlation scenario $\mathcal{E}_{S,\Lambda }$ is called \emph{%
nonsignaling} if, for any two joint measurements $(s_{1},...,s_{N})$ and $%
(s_{1}^{\prime },...,s_{N}^{\prime })$ with common settings $%
s_{n_{1}},...,s_{n_{_{M}}}$ at some $1\leq n_{1}<$ $...<n_{M}\leq N$ sites,
the marginal probability distributions of $P_{(s_{1},...,s_{N})}^{(\mathcal{E%
}_{S,\Lambda })}$ and $P_{(s_{1}^{\prime },...,s_{N}^{\prime })}^{(\mathcal{E%
}_{S,\Lambda })}$, describing measurements at these sites, coincide. For
details on the mathematical specification of nonsignaling and the EPR
locality, see section 3 in \cite{5}.

For a correlation scenario $\mathcal{E}_{S,\Lambda },$ consider a linear
combination 
\begin{eqnarray}
\mathcal{B}_{\Phi _{S,\Lambda }}^{(\mathcal{E}_{S,\Lambda })}
&=&\sum_{s_{1},...,s_{_{N}}}\left\langle f_{(s_{1},...,s_{N})}(\lambda
_{1},\ldots ,\lambda _{N})\right\rangle _{\mathcal{E}_{S,\Lambda }},
\label{1} \\
\Phi _{S,\Lambda } &=&\{f_{(s_{1},...,s_{N})}:\Lambda \rightarrow \mathbb{R}%
\mid \text{ }s_{n}=1,...,S_{n},\text{ }n=1,...,N\},  \notag
\end{eqnarray}%
of averages 
\begin{eqnarray}
&&\left\langle f_{(s_{1},...,s_{N})}(\lambda _{1},\ldots ,\lambda
_{N})\right\rangle _{\mathcal{E}_{S,\Lambda }}  \label{2} \\
&=&\int\limits_{\Lambda }f_{(s_{1},...,s_{N})}(\lambda _{1},\ldots ,\lambda
_{N})P_{(s_{1},...,s_{N})}^{(\mathcal{E}_{S,\Lambda })}\left( \mathrm{d}%
\lambda _{1}\times \cdots \times \mathrm{d}\lambda _{N}\right)  \notag
\end{eqnarray}%
of the most general form, specified for each joint measurement $%
(s_{1},...,s_{N})$ by a bounded real-valued function $f_{(s_{1},...,s_{N})}(%
\lambda _{1},\ldots ,\lambda _{N})$ of measurement outcomes $\left( \lambda
_{1},\ldots ,\lambda _{N}\right) \in \Lambda $ at all $N$ sites.

Depending on a choice of a function $f_{(s_{1},...,s_{N})}$, an average (\ref%
{2}) may refer either to the joint probability of events at $M\leq N$ sites
or, for example, in case of real-valued outcomes at each $n$-th site, to the
mean value 
\begin{equation}
{\Large \langle }\lambda _{1}^{(s_{1})}\cdot \ldots \cdot \lambda
_{n_{M}}^{(s_{n_{M}})}{\Large \rangle }_{\mathcal{E}_{S,\Lambda
}}=\int\limits_{\Lambda }\lambda _{1}\cdot \ldots \cdot \lambda
_{n_{M}}P_{(s_{1},...,s_{N})}^{(\mathcal{E}_{S,\Lambda })}\left( \mathrm{d}%
\lambda _{1}\times \cdots \times \mathrm{d}\lambda _{N}\right)  \label{3}
\end{equation}%
(expectation) of the product of outcomes observed at $M\leq N$ sites under a
joint measurement $(s_{1},...,s_{N})$. In quantum information, the average (%
\ref{3}) is referred to as a correlation function. For $M=N,$ a correlation
function is called full.

The probabilistic description of an arbitrary correlation scenario $\mathcal{%
E}_{S,\Lambda }$ admits\footnote{%
For the main statements on the LHV modelling of a general multipartite
correlation scenario, see section 4 in \cite{5}.} \emph{a LHV (local hidden
variable) probability model} if each of its joint probability distributions%
\begin{equation}
\left \{ P_{(s_{1},...,s_{N})}^{(\mathcal{E}_{S,\Lambda })},\text{ }%
s_{1}=1,...,S_{n},...,s_{N}=1,...,S_{N}\right \}  \label{4}
\end{equation}%
admits the representation%
\begin{eqnarray}
&&P_{(s_{1},...,s_{N})}^{(\mathcal{E}_{S,\Lambda })}\left( \mathrm{d}\lambda
_{1}\times \cdots \times \mathrm{d}\lambda _{N}\right)  \label{5} \\
&=&\dint \limits_{\Omega }P_{1,s_{1}}(\mathrm{d}\lambda _{1}|\omega )\cdot
\ldots \cdot P_{N,s_{N}}(\mathrm{d}\lambda _{N}|\omega )\text{ }\nu _{%
\mathcal{E}_{S,\Lambda }}^{lhv}(\mathrm{d}\omega )  \notag
\end{eqnarray}%
via a single probability distribution $\nu _{\mathcal{E}_{S,\Lambda }}^{lhv}$
of some variables $\omega \in \Omega $ and conditional probability
distributions $P_{n,s_{n}}(\mathrm{\cdot }|\omega )$ of outcomes $\lambda
_{n}$ at $n$-th site, referred to as "local" in the sense that each $%
P_{n,s_{n}}$ depends only on a measurement setting $s_{_{n}}$ at $n$-th site.

In quantum theory, variables $\omega \in \Omega $ are generally referred to
as\footnote{%
This terminology formed historically, see Introduction in \cite{5}.} "hidden
variables" (HV) -- this and "locality" of distributions $P_{n,s_{n}}$
explains the title "LHV" of this model.

Note that, though, in the general LHV representation (\ref{5}), each
distribution $P_{n,s_{n}}(\cdot |\omega )$ depends only on a measurement
setting $s_{_{n}}$ at $n$-th site, a probability distribution $\nu _{%
\mathcal{E}_{S,\Lambda }}^{lhv}$ of variables $\omega ,$ which has \emph{a
simulation character, }may, in general, depend via the subscript $\mathcal{E}%
_{S,\Lambda }$ on measurement settings at all (or some) sites. This is, for
example, the case in quantum LHV models considered in section 5 of \cite{5}.

If, in addition to representation (\ref{5}), some distributions $%
P_{n,s_{n}}(\cdot |\omega )$ corresponding to different sites are correlated
(via $\omega $), then we refer \cite{5} to such an LHV model as \emph{%
conditional}. The LHV model considered by Bell in \cite{1} represents an
example of a conditional LHV model.

From representation (\ref{5}) it follows that each LHV correlation scenario
is nonsignaling, though not vice versa.

Let a correlation scenario $\mathcal{E}_{S,\Lambda }$ admit a LHV model.
Then a linear combination (\ref{1}) of its averages satisfies the tight LHV
constraints \cite{23}%
\begin{equation}
\mathcal{B}_{\Phi _{S,\Lambda }}^{\inf }\leq \mathcal{B}_{\Phi _{S,\Lambda
}}^{(\mathcal{E}_{S,\Lambda })}{\Large |}_{_{lhv}}\leq \mathcal{B}_{\Phi
_{S,\Lambda }}^{\sup }  \label{6}
\end{equation}%
with the LHV constants%
\begin{eqnarray}
\mathcal{B}_{\Phi _{S,\Lambda }}^{\sup } &=&\sup_{\lambda _{n}^{(s_{n})}\in
\Lambda _{n},\forall s_{n},\forall n}\text{ }%
\sum_{s_{1},...,s_{_{N}}}f_{(s_{1},...,s_{N})}(\lambda _{1}^{(s_{1})},\ldots
,\lambda _{N}^{(s_{N})}),  \label{7} \\
\mathcal{B}_{\Phi _{S,\Lambda }}^{\inf } &=&\inf_{\lambda _{n}^{(s_{n})}\in
\Lambda _{n},\forall s_{n},\forall n}\text{ }%
\sum_{s_{1},...,s_{_{N}}}f_{(s_{1},...,s_{N})}(\lambda _{1}^{(s_{1})},\ldots
,\lambda _{N}^{(s_{N})}).  \label{8}
\end{eqnarray}

If a correlation scenario $\mathcal{E}_{S,\Lambda }$ admits a \emph{%
conditional} LHV model, then a linear combination $\mathcal{B}_{\Phi
_{S,\Lambda }}^{(\mathcal{E}_{S,\Lambda })}$ of its averages satisfies not
only unconditional LHV constraints (\ref{6}) but also their conditional
version where the LHV constants $\mathcal{B}_{\Phi _{S,\Lambda }}^{\sup
}|_{cond}$ and $\mathcal{B}_{\Phi _{S,\Lambda }}^{\inf }|_{cond}$ are
defined similarly to (\ref{7}), (\ref{8}) but \emph{via} \emph{conditional
supremum and infimum.}

Some of the LHV constraints (\ref{6}) may be fulfilled for a wider (than
LHV) class of correlation scenarios. This is, for example, the case for the
LHV\ constraints on joint probabilities following explicitly from
nonsignaling of probability distributions. Moreover, some of constraints (%
\ref{6}) may be simply trivial, i. e. fulfilled for all correlation
scenarios, not necessarily nonsignaling.

\emph{Each of the tight linear LHV constraints (\ref{6}) that may be
violated under a non-LHV scenario is referred to as a Bell (or Bell-type)
inequality. }

If outcomes at each $n$-th site are real-valued, let $\lambda _{n}\in 
\widetilde{\Lambda }_{n}=[-1,1],$ and we specify (\ref{6}) for the full
correlation functions, that is, for the collection $\widetilde{\Phi }_{S,%
\widetilde{\Lambda }}$ of functions $f_{(s_{1},...,s_{N})}(\lambda _{1}\cdot
\ldots \cdot \lambda _{N})=$ $\alpha _{(s_{1},...,s_{N})}\lambda _{1}\cdot
\ldots \cdot \lambda _{N},$ then $\mathcal{B}_{\widetilde{\Phi }_{S,%
\widetilde{\Lambda }}}^{\sup }=-\mathcal{B}_{\widetilde{\Phi }_{S,\widetilde{%
\Lambda }}}^{\inf }$ and a Bell inequality (\ref{6}) on the full correlation
functions takes the form \cite{23}:%
\begin{eqnarray}
&&{\LARGE |}\sum_{s_{1},...,s_{_{N}}}\alpha _{(s_{1},...,s_{N})}\left\langle
\lambda _{1}^{(s_{1})}\cdot \ldots \cdot \lambda _{N}^{(s_{N})}\right\rangle
_{lhv}{\LARGE |}  \label{10} \\
&\leq &\max_{\eta _{n}\in \lbrack -1,1]^{S_{n}},\forall n}\left\vert
F_{\alpha }(\eta _{1},...,\eta _{N})\right\vert =\max_{\eta _{n}\in
\{-1,1\}^{S_{n}},\forall n}\left\vert F_{\alpha }(\eta _{1},...,\eta
_{N})\right\vert ,  \notag
\end{eqnarray}%
where%
\begin{equation}
F_{\alpha }(\eta _{1},...,\eta _{N})=\sum_{s_{1},...,s_{_{N}}}\alpha
_{(s_{1},...,s_{N})}\eta _{1}^{(s_{1})}\cdot \ldots \cdot \eta _{N}^{(s_{N})}
\label{11}
\end{equation}%
is the $N$-linear form of $S_{n}$-dimensional real-valued vectors $\eta _{n}$
$=$ $(\eta _{n}^{(1)},...,$ $\eta _{n}^{(S_{n})})$ $\in \lbrack
-1,1]^{S_{n}}.$

Note that, the value of the maximum in the right-hand side of (\ref{10})
does not depend on a number of measurement outcomes at each site and is
determined only by the extreme values $\pm 1$ of these outcomes. Therefore, 
\emph{the form of each correlation Bell inequality (\ref{10}) does not
depend on a spectral type of outcomes at each site, in particular, on their
number.} This observation is rather essential since, in the polytope
approach, the classification of correlation Bell inequalities essentially
depends on a number of measurement outcomes at each site, see, for example,
in \cite{18}.

The specification of the general representation (\ref{6})\ for Bell
inequalities on joint probabilities is derived quite similarly and is
presented by Eq. (39) in \cite{23}, incorporating in a unique manner all
Bell inequalities on joint probabilities derived in the literature via the
other approaches, for example, the Bell inequalities in \cite{28}.

\section{The LqHV modelling}

As it is well known since the seminal paper \cite{1} of Bell, the
probabilistic description of an arbitrary quantum correlation scenario
cannot be reproduced via a LHV model.

However, as we proved in \cite{6, 21}, the probabilistic description of
every quantum correlation scenario, more generally, every nonsignaling
scenario, can be correctly reproduced via \emph{a LqHV} $\emph{(}$\emph{%
local quasi hidden variable) probability model }-- the notion\emph{\ }%
introduced for an arbitrary correlation scenario in \cite{6}.\ Moreover, all
quantum correlation scenarios on an $N$-partite state with projective
quantum measurements at each site admit \cite{8, 22} a \emph{single} LqHV
model.

In a LqHV model, all scenario joint probability distributions (\ref{4})
admit the representation 
\begin{eqnarray}
&&P_{(s_{1},...,s_{N})}^{(\mathcal{E}_{S,\Lambda })}\left( \mathrm{d}\lambda
_{1}\times \cdots \times \mathrm{d}\lambda _{N}\right)  \label{12} \\
&=&\dint \limits_{\Omega }P_{1,s_{1}}(\mathrm{d}\lambda _{1}|\omega )\cdot
\ldots \cdot P_{N,s_{N}}(\mathrm{d}\lambda _{N}|\omega )\text{ }\mu _{%
\mathcal{E}_{S,\Lambda }}^{lqhv}(\mathrm{d}\omega ),  \notag
\end{eqnarray}%
which is quite similar by its form to the LHV representation (\ref{5}) with
only one difference -- in (\ref{12}), a normalized distribution $\mu _{%
\mathcal{E}_{S,\Lambda }}^{lqhv}$ of variables $\omega \in \Omega $ is
real-valued and does not need to be positive.

Therefore, a LHV model (\ref{5}) constitutes a particular case of a LqHV
probability model\emph{\ }(\ref{12}) whenever a distribution $\mu _{\mathcal{%
E}_{S,\Lambda }}^{lqhv}$ is positive. Also, the affine model \cite{29} for a
family of nonsignaling probability distributions constitutes a LqHV model of
a particular type.

Clearly, a LqHV model (\ref{12}) is "local" in the same sense as it was
meant by Bell \cite{1} for the LHV representation (\ref{5}). The term
"quasi" in its title "LqHV" refers only to "hidden variables" (HV),
specifically, to a possible nonpositivity of a distribution $\mu _{\mathcal{E%
}_{S,\Lambda }}^{lqhv}$ of these variables.

For an arbitrary correlation scenario $\mathcal{E}_{S,\Lambda }$, the
following statements are equivalent \cite{21}:

\begin{itemize}
\item the probabilistic description of a correlation scenario admits a LqHV
model (\ref{12});

\item a correlation scenario is nonsignaling;

\item there is a real-valued distribution 
\begin{equation}
\tau _{\mathcal{E}_{S,\Lambda }}^{lqhv}\left( \mathrm{d}\lambda
_{1}^{(1)}\times \cdots \times \mathrm{d}\lambda _{1}^{(S_{1})}\times \cdots
\times \mathrm{d}\lambda _{N}^{(1)}\times \cdots \times \mathrm{d}\lambda
_{N}^{(S_{N})}\right)  \label{13}
\end{equation}%
of all scenario outcomes, returning each scenario joint probability
distribution $P_{(s_{1},...,s_{N})}^{(\mathcal{E}_{S,\Lambda })}$ as the
corresponding marginal.
\end{itemize}

Note that a nonsignaling correlation scenario, which we further specify by $%
\mathcal{E}_{S,\Lambda }^{ns},$ may admit a variety of LqHV models (\ref{12}%
).

\section{Nonsignaling analogs and Bell's nonlocality}

Let us now construct analogs of the LHV\ constraints (\ref{6}) for a general
nonsignaling case. Substituting into averages (\ref{2}) of a linear
combination (\ref{1}) the LqHV representation (\ref{12}) for joint
probability distributions $P_{(s_{1},...,s_{N})}^{(\mathcal{E}_{S,\Lambda
}^{ns})},$ recalling for a normalized real-valued distribution $\mu $ the
Jordan decomposition\footnote{%
For this decomposition see, for example, section 3 in \cite{6}.} via
positive distributions $\mu ^{(\pm )}:$%
\begin{equation}
\mu =\mu ^{(+)}-\mu ^{(-)},\text{ \ \ }\mu ^{(+)}(\Omega )-\mu ^{(-)}(\Omega
)=1,  \label{14.1}
\end{equation}%
and minimizing over all possible LqHV models for $\mathcal{E}_{S,\Lambda
}^{ns}$, we come to the following analogs of constraints (\ref{6}) for a
nonsignaling scenario $\mathcal{E}_{S,\Lambda }^{ns}$:%
\begin{align}
& \mathcal{B}_{\Phi _{S,\Lambda }}^{\inf }-\frac{\gamma _{\mathcal{E}%
_{S,\Lambda }^{ns}}-1}{2}(\mathcal{B}_{\Phi _{S,\Lambda }}^{\sup }-\mathcal{B%
}_{\Phi _{S,\Lambda }}^{\inf })  \label{14} \\
& \leq \mathcal{B}_{\Phi _{S,\Lambda }}^{(\mathcal{E}_{S,\Lambda }^{ns})} 
\notag \\
& \leq \mathcal{B}_{\Phi _{S,\Lambda }}^{\sup }+\frac{\gamma _{\mathcal{E}%
_{S,\Lambda }^{ns}}-1}{2}(\mathcal{B}_{\Phi _{S,\Lambda }}^{\sup }-\mathcal{B%
}_{\Phi _{S,\Lambda }}^{\inf }),  \notag
\end{align}%
where the parameter $\gamma _{\mathcal{E}_{S,\Lambda }^{ns}}$ has the form 
\begin{eqnarray}
\gamma _{\mathcal{E}_{S,\Lambda }^{ns}} &=&\inf_{\mu _{_{\mathcal{E}%
_{S,\Lambda }^{ns}}}^{lqhv}}\left \Vert \mu _{_{\mathcal{E}_{S,\Lambda
}^{ns}}}^{lqhv}\right \Vert _{var}\geq 1,  \label{16} \\
||\mu _{_{\mathcal{E}_{S,\Lambda }^{ns}}}^{lqhv}||_{var} &=&(\mu _{_{%
\mathcal{E}_{S,\Lambda }^{ns}}}^{lqhv})^{(+)}(\Omega )+(\mu _{_{\mathcal{E}%
_{S,\Lambda }^{ns}}}^{lqhv})^{(-)}(\Omega )\geq 1,  \notag
\end{eqnarray}%
with infimum taken over all possible LqHV models (\ref{12}) for a scenario $%
\mathcal{E}_{S,\Lambda }^{ns}$ and notation $\left \Vert \mu \right \Vert
_{var}$ meaning the total variation norm of a real-valued distribution $\mu $%
. For a normalized real-valued distribution $\mu ,$ this norm $\left \Vert
\mu \right \Vert _{var}\geq 1,$ with $\left \Vert \mu _{var}\right \Vert =1$
if and only if $\mu $ is a probability distribution. For a discrete
distribution, the total variation norm reduces to the sum of all its
absolute values.

From (\ref{16}) it follows that the parameter $\gamma _{\mathcal{E}%
_{S,\Lambda }^{ns}},$ specifying in (\ref{14}) violation of a Bell
inequality under a nonsignaling scenario $\mathcal{E}_{S,\Lambda }^{ns}$,
does not depend on a form of this inequality and $\gamma _{\mathcal{E}%
_{S,\Lambda }^{ns}}=1$ ( no violation) if and only if a nonsignaling
scenario $\mathcal{E}_{S,\Lambda }^{ns}$ is an LHV\ one. Moreover, it is
easy to prove, quite similarly to our proof of Lemma 3 in \cite{6}, that the
parameter $\gamma _{\mathcal{E}_{S,\Lambda }^{ns}},$ given by (\ref{16}), is
otherwise expressed as $\gamma $ 
\begin{eqnarray}
\gamma _{\mathcal{E}_{S,\Lambda }^{ns}} &=&\sup_{\Phi _{S,\Lambda }}\frac{1}{%
\mathcal{B}_{\Phi _{S,\Lambda }}^{lhv}}\left \vert \mathcal{B}_{\Phi
_{S,\Lambda }}^{(\mathcal{E}_{S,\Lambda }^{ns})}\right \vert ,  \label{17} \\
\mathcal{B}_{\Phi _{S,\Lambda }}^{lhv} &=&\max {\Large \{}|\mathcal{B}_{\Phi
_{S,\Lambda }}^{\inf }|,|\mathcal{B}_{\Phi _{S,\Lambda }}^{\sup }|{\large \},%
}  \notag
\end{eqnarray}%
that is, constitutes the maximal violation under a nonsignaling scenario $%
\mathcal{E}_{S,\Lambda }^{ns}$ of \emph{all} \emph{general} Bell
inequalities for $S_{n}$ settings and outcomes $\lambda _{n}\in \Lambda _{n}$
at each $n$-th site.

In order to construct analogs of Bell inequalities for an arbitrary class $%
\mathfrak{G}_{ns}$ of nonsignaling scenarios $\mathcal{E}_{S,\Lambda }^{ns}$
with $S_{1},...,S_{N}$ settings and outcome sets $\Lambda _{1},...,\Lambda
_{N}$ at $N$ sites, for example, for all quantum correlation scenarios $%
\mathcal{E}_{S,\Lambda }^{\rho }$ on a state $\rho $ or for all possible
nonsignaling scenarios $\mathcal{E}_{S,\Lambda }^{ns},$ we maximize (\ref{14}%
) over all scenarios $\mathcal{E}_{S,\Lambda }^{ns}\in \mathfrak{G}_{ns}$
and come to the following analogs of Bell inequalities (\ref{6}) in a
general nonsignaling case:%
\begin{align}
& \mathcal{B}_{\Phi _{S,\Lambda }}^{\inf }-\frac{\mathrm{\Upsilon }%
_{S,\Lambda }^{\mathfrak{G}_{ns}}-1}{2}(\mathcal{B}_{\Phi _{S,\Lambda
}}^{\sup }-\mathcal{B}_{\Phi _{S,\Lambda }}^{\inf })  \label{18} \\
& \leq \mathcal{B}_{\Phi _{S,\Lambda }}^{(\mathcal{E}_{S,\Lambda }^{ns})}%
{\LARGE |}_{\mathfrak{G}_{ns}}  \notag \\
& \leq \mathcal{B}_{\Phi _{S,\Lambda }}^{\sup }+\frac{\mathrm{\Upsilon }%
_{S,\Lambda }^{\mathfrak{G}_{ns}}-1}{2}(\mathcal{B}_{\Phi _{S,\Lambda
}}^{\sup }-\mathcal{B}_{\Phi _{S,\Lambda }}^{\inf }).  \notag
\end{align}%
In (\ref{18}), the LqHV parameter $\mathrm{\Upsilon }_{S,\Lambda }^{%
\mathfrak{G}_{ns}}$ is given by%
\begin{equation}
\mathrm{\Upsilon }_{S,\Lambda }^{\mathfrak{G}_{ns}}=\sup_{\mathcal{E}%
_{S,\Lambda }^{ns}\in \mathfrak{G}_{ns}}\gamma _{\mathcal{E}_{S,\Lambda
}^{ns}}=\sup_{\mathcal{E}_{S,\Lambda }^{ns}\in \mathfrak{G}_{ns}}\inf_{\mu
_{_{\mathcal{E}_{S,\Lambda }^{ns}}}^{lqhv}}\left \Vert \mu _{_{\mathcal{E}%
_{S,\Lambda }^{ns}}}^{lqhv}\right \Vert _{var}\geq 1  \label{20}
\end{equation}%
and, in view of (\ref{17}), constitutes the maximal violation%
\begin{equation}
\mathrm{\Upsilon }_{S,\Lambda }^{\mathfrak{G}_{ns}}=\sup_{\mathcal{E}%
_{S,\Lambda }^{ns}\in \mathfrak{G}_{ns},\text{ }\Phi _{S,\Lambda }}\frac{1}{%
\mathcal{B}_{\Phi _{S,\Lambda }}^{lhv}}\left \vert \mathcal{B}_{\Phi
_{S,\Lambda }}^{(\mathcal{E}_{S,\Lambda }^{ns})}\right \vert \geq 1
\label{21}
\end{equation}%
under nonsignaling scenarios $\mathcal{E}_{S,\Lambda }^{ns}\in \mathfrak{G}%
_{ns}$ of general Bell inequalities for $S_{n}$ settings and outcomes $%
\lambda _{n}\in \Lambda _{n}$ at each $n$-th site. Clearly, $\mathrm{%
\Upsilon }_{S,\Lambda }^{\mathfrak{G}_{ns}}=1$ if and only if $\mathfrak{G}%
_{ns}$ is a class of nonsignaling scenarios, each admitting a LHV model.

As an example, consider the nonsignaling analogs (\ref{18}) for the
Clauser-Horne (CH) inequalities\footnote{%
For these Bell inequalities, see, for example, subsection 3.2 in \cite{23}.}
on joint probabilities%
\begin{equation}
-1\leq \mathcal{B}_{CH}{\LARGE |}_{lhv}\leq 0.  \label{22.1}
\end{equation}%
These Bell inequalities correspond to the bipartite case with two settings
and two outcomes per site -- in notation of \cite{28}, this is the $(2222)$
case.

From (\ref{18}) and (\ref{22.1}) it follows that, for an arbitrary class $%
\mathfrak{G}_{ns}$ of nonsignaling scenarios $(2222)$, the analogs of the CH
inequalities take the form:%
\begin{equation}
-\frac{\mathrm{\Upsilon }_{2222}^{\mathfrak{G}_{ns}}+1}{2}\leq \mathcal{B}%
_{CH}{\LARGE |}_{\mathfrak{G}_{ns}}\leq \frac{\mathrm{\Upsilon }_{2222}^{%
\mathfrak{G}_{ns}}-1}{2}.  \label{22}
\end{equation}%
For example, for all quantum correlation scenarios $(2222)$, the maximal
quantum Bell violation $\mathrm{\Upsilon }_{2222}^{\text{quant}}=\sqrt{2}$
and inequalities (\ref{22}) reduce to the well-known quantum analogs of the
CH inequalities.

From (\ref{18}), (\ref{20}), (\ref{21}) it follows that, for a $\mathfrak{G}%
_{ns}$-class of nonsignaling scenarios with $S_{1},...,S_{N}$ settings but
arbitrary outcome sets $\Lambda _{1},...,\Lambda _{N}$ at $N$ sites, the
parameter%
\begin{equation}
\mathrm{\Upsilon }_{S_{1}\times \cdots \times S_{N}}^{\mathfrak{G}%
_{ns}}=\sup_{\Lambda ,\text{ }\mathcal{E}_{S,\Lambda }^{ns}\in \mathfrak{G}%
_{ns}}\left( \inf_{\mu _{_{\mathcal{E}_{S,\Lambda }^{ns}}}^{lqhv}}\left
\Vert \mu _{_{\mathcal{E}_{S,\Lambda }^{ns}}}^{lqhv}\right \Vert
_{var}\right) \geq 1  \label{23}
\end{equation}%
quantifies the $S_{1}\times \cdots \times S_{N}$-setting Bell's nonlocality
whereas the parameter%
\begin{equation}
\mathrm{\Upsilon }_{\mathfrak{G}_{ns}}=\sup_{S_{1},..,S_{N}}\mathrm{\Upsilon 
}_{S_{1}\times \cdots \times S_{N}}^{\mathfrak{G}_{ns}}\geq 1  \label{24}
\end{equation}%
-- the full Bell's nonlocality. Locality, the $S_{1}\times \cdots \times
S_{N}$-setting and full, corresponds to $\mathrm{\Upsilon }_{S_{1}\times
\cdots \times S_{N}}^{\mathfrak{G}_{ns}}=1$ and $\mathrm{\Upsilon }_{%
\mathfrak{G}_{ns}}=1,$ respectively.

Note that if $\mathrm{\Upsilon }_{S_{1}\times \cdots \times S_{N}}^{%
\mathfrak{G}_{ns}}=1$ for some integers $S_{1},...,S_{N}\geq 1,$ then $%
\mathrm{\Upsilon }_{S_{1}^{\prime }\times \cdots \times S_{N}^{\prime }}^{%
\mathfrak{G}_{ns}}=1$ for all positive integers $S_{1}^{\prime }\leq S_{1},$ 
$...,$ $S_{N}^{\prime }\leq S_{N}$.

\section{Quantum nonlocality}

Consider now the nonlocality parameters $\mathrm{\Upsilon }_{S_{1}\times
\cdots \times S_{N}}^{\mathfrak{G}_{ns}}$ and $\mathrm{\Upsilon }_{\mathfrak{%
G}_{ns}}$ for a particular class $\mathfrak{G}_{ns}$ of nonsignaling
scenarios -- quantum correlation scenarios performed on a state $\rho $ on a
Hilbert space $\mathcal{H}_{1}\otimes \cdots \otimes \mathcal{H}_{N}.$

For this class of nonsignaling scenarios, we shortly denote 
\begin{equation}
\mathrm{\Upsilon }_{S_{1}\times \cdots \times S_{N}}^{\mathfrak{G}%
_{ns}}\rightarrow \mathrm{\Upsilon }_{S_{1}\times \cdots \times
S_{N}}^{(\rho )},\text{ \ \ \ \ }\mathrm{\Upsilon }_{\mathfrak{G}%
_{ns}}\rightarrow \mathrm{\Upsilon }_{\rho },  \label{25}
\end{equation}%
and, as it is generally accepted, refer to Bell's nonlocality as \emph{%
quantum nonlocality}.

According to (\ref{23}), (\ref{24}), an $N$-partite quantum state $\rho $\
is:

\begin{itemize}
\item the $S_{1}\times \cdots \times S_{N}$-setting nonlocal if and only if $%
\Upsilon _{S_{1}\times \cdots \times S_{N}}^{(\rho )}>1;$

\item fully nonlocal if and only if $\Upsilon _{\rho }>1$\emph{.}
\end{itemize}

Let us now evaluate the quantum nonlocality parameters $\Upsilon
_{S_{1}\times \cdots \times S_{N}}^{(\rho )}$ and $\Upsilon _{\rho }$.

We recall that, \emph{for every quantum state} $\rho $ on $\mathcal{H}%
_{1}\otimes \cdots \otimes \mathcal{H}_{N}$ and arbitrary positive integers $%
S_{1},\ldots ,S_{N}\geq 1$, \emph{there exists} \cite{6, 14, 15} an $%
S_{1}\times \cdots \times S_{N}$-setting source operator $T_{S_{1}\times
\cdots \times S_{N}}^{(\rho )}$ -- that is, a self-adjoint trace class
dilation of a state $\rho $ to the space 
\begin{equation}
(\mathcal{H}_{1})^{\otimes S_{1}}\otimes \cdots \otimes (\mathcal{H}%
_{N})^{\otimes S_{N}}.  \label{26}
\end{equation}%
Clearly, $T_{1\times \cdots \times 1}^{(\rho )}=\rho $ and $\mathrm{tr}%
[T_{S_{1}\times \cdots \times S_{N}}^{(\rho )}]=1.$

In view of the analytical upper bound (53) in \cite{6}, we have: 
\begin{eqnarray}
\mathrm{\Upsilon }_{S_{1}\times \cdots \times S_{N}}^{(\rho )} &\leq
&\inf_{T_{S_{1}\times \cdots \times \underset{\overset{\uparrow }{n}}{1}%
\times \cdots \times S_{N}}^{(\rho )},\text{ }\forall n}||T_{S_{1}\times
\cdots \times \underset{\overset{\uparrow }{n}}{1}\times \cdots \times
S_{N}}^{(\rho )}||_{cov},  \label{27} \\
\mathrm{\Upsilon }_{\rho } &\leq &\sup_{S_{1},...,S_{N}}\left(
\inf_{T_{S_{1}\times \cdots \times \underset{\overset{\uparrow }{n}}{1}%
\times \cdots \times S_{N}}^{(\rho )},\text{ }\forall n}||T_{S_{1}\times
\cdots \times \underset{\overset{\uparrow }{n}}{1}\times \cdots \times
S_{N}}^{(\rho )}\text{ }||_{cov}\right) ,  \notag
\end{eqnarray}%
where infimum is taken over all source operators $T_{S_{1}\times \cdots
\times \underset{\overset{\uparrow }{n}}{1}\times \cdots \times
S_{N}}^{(\rho )}$ with only one setting at some $n$-th site and over all
sites $n=1,...,N,$ and notation $\left \Vert \cdot \right \Vert _{cov}$
means \emph{the covering norm} -- a new norm introduced for self-adjoint
trace class operators by relation (11) in \cite{6}.

For every self-adjoint trace class operator $W$ on a tensor product Hilbert
space $\mathcal{G}_{1}\otimes \mathcal{\cdots }\otimes \mathcal{G}_{m},$ the
covering norm satisfies \cite{6} the relation%
\begin{equation}
\left \vert \mathrm{tr}\left[ W\right] \right \vert \leq \left \Vert W\right
\Vert _{cov}\leq \left \Vert W\right \Vert _{1},  \label{29}
\end{equation}%
where $\left \Vert \cdot \right \Vert _{1}$ is the trace norm and the
equality $\left \Vert W\right \Vert _{cov}=\left \vert \mathrm{tr}\left[ W%
\right] \right \vert $ is true if a self-adjoint trace class operator $W$ is
tensor positive \cite{6}, that is, 
\begin{equation}
\mathrm{tr}\left[ W\{X_{1}\otimes \cdots \otimes X_{m}\} \right] \geq 0
\label{30}
\end{equation}%
for all positive bounded operators $X_{j}$ on $\mathcal{G}_{j},$ $j=1,...,m$.

Since, for every source operator, $\mathrm{tr}[T_{S_{1}\times \cdots \times
S_{N}}^{(\rho )}]=1,$ from (\ref{29}) it follows that $||T_{S_{1}\times
\cdots \times S_{N}}^{(\rho )}||_{cov}\geq 1,$ $\forall T_{S_{1}\times
\cdots \times S_{N}}^{(\rho )},$ and $||T_{S_{1}\times \cdots \times
S_{N}}^{(\rho )}||_{cov}$ $=1$ if $T_{S_{1}\times \cdots \times
S_{N}}^{(\rho )}$ is tensor positive. This and the analytical upper bounds (%
\ref{27}) imply.

\emph{If, for an }$N$\emph{-partite quantum state }$\rho $ \emph{and
arbitrary integers} $S_{1},...,S_{N}$ $\geq 1,$\emph{\ there exists a tensor
positive source operator }$T_{S_{1}\times \cdots \times \underset{\overset{%
\uparrow }{n}}{1}\times \cdots \times S_{N}}^{(\rho )}$\emph{\ for some }$%
n=1,...,N$\emph{, then }$\mathrm{\Upsilon }_{S_{1}\times \cdots \times
S_{N}}^{(\rho )}=1\ $\emph{and} \emph{this }$N$\emph{-partite} \emph{state
is the }$S_{1}\times \cdots \times S_{N}$\emph{-setting local, that is,
satisfies all general }$S_{1}^{\prime }\times \cdots \times S_{N}^{\prime }$%
\emph{-setting Bell inequalities with }$S_{1}^{\prime }\leq
S_{1},...,S_{N}^{\prime }\leq S_{N}$\emph{\ measurement settings at N sites. 
}

\emph{If, for an }$N$\emph{-partite quantum state }$\rho ,$\emph{\ tensor
positive source operators }$T_{S_{1}\times \cdots \times \underset{\overset{%
\uparrow }{n}}{1}\times \cdots \times S_{N}}^{(\rho )}$\emph{\ for some
arbitrary }$n=1,...,N$\emph{, exist for all integers} $S_{1},...,S_{N}\geq
1, $ \emph{then} $\mathrm{\Upsilon }_{\rho }=1$ $\emph{and}$ \emph{this }$N$%
\emph{-partite state }$\rho $ \emph{is (fully) local, that is, satisfies all
general Bell inequalities.\smallskip }

For an $N$-qudit quantum state $\rho _{d,N}$ on $(\mathcal{H})^{\otimes N},$ 
$d=\dim \mathcal{H}<\infty ,$ let us now evaluate the analytical upper
bounds (\ref{27}) in $d,$ $S$ and $N$.

From (\ref{21}) and our results in \cite{6, 8, 20} it follows that the
quantum nonlocality parameters $\mathrm{\Upsilon }_{\rho _{d,N}}$ and $%
\Upsilon _{S\times \cdots \times S}^{(\rho _{d,N})}$ admit the upper bounds 
\begin{eqnarray}
\Upsilon _{S\times \cdots \times S}^{(\rho _{d,N})} &\leq &\left( 2\min
\{d,S\}-1\right) ^{N-1},  \label{32} \\
\mathrm{\Upsilon }_{\rho _{d,N}} &\leq &(2d-1)^{N-1},  \notag
\end{eqnarray}%
under all generalized $N$-partite quantum measurements and the more specific
upper bounds 
\begin{eqnarray}
\Upsilon _{2\times \cdots \times 2}^{(\rho _{d,N})} &\leq &\min {\Large \{}%
d^{\frac{N-1}{2}},\text{ }3^{N-1}{\Large \}},\text{ \ \ \ \ for \ }S=2,
\label{34} \\
\Upsilon _{S\times \cdots \times S}^{(\rho _{d,N})} &\leq &\min {\Large \{}%
d^{\frac{S(N-1)}{2}},\text{ }\left( 2\min \{d,S\}-1\right) ^{N-1}{\Large \}},%
\text{ \ \ for \ }S\geq 3,  \notag
\end{eqnarray}%
under projective $N$-partite quantum measurements.

These upper bounds are attainable. For $N=d=S=2,$ the upper bound (\ref{34})
is attained on the CHSH inequality. For $d=S=2,$ $N\geq 3,$ it is attained
on the Mermin--Klyshko inequality.

For the analysis of attainability of (\ref{34}) beyond the two-qubit case,
let us consider the quantum analogs of the Zohren-Gill (ZG) inequalities 
\cite{24} on joint probabilities:%
\begin{equation}
1\leq \mathcal{B}_{ZG}|_{_{_{_{lhv}}}}\leq 2.  \label{35}
\end{equation}%
These Bell inequalities correspond to the bipartite case with two settings $%
(N=S=2)$ and $d$ outcomes at each site. For this case, the critical value in
the upper bound (\ref{34}) is equal to $\min \{ \sqrt{d},3\}$ and by (\ref%
{18}) this critical value leads to the following quantum analogs of the ZG
inequalities under projective bipartite quantum measurements:%
\begin{eqnarray}
\frac{3-\sqrt{d}}{2} &\leq &\mathcal{B}_{ZG}{\Large |}_{_{_{\rho
_{d,N}}}}\leq \frac{3+\sqrt{d}}{2},\text{ \ \ \ for \ }d\leq 9,  \label{36}
\\
0 &\leq &\mathcal{B}_{ZG}{\Large |}_{_{_{\rho _{d,N}}}}\leq 3,\text{ \ \ \
for \ }d\geq 9.  \label{37}
\end{eqnarray}%
The left-hand side inequality in (\ref{37})\emph{\ just coincides} with
inequality (8) in \cite{24}, which was conjectured by Zohren and Gill (in
view of their numerical results) as the quantum analog of the Bell
inequality (\ref{35}) for an infinite dimensional case.

\section{Conceptual issues}

In the present paper, we have analyzed violation of Bell inequalities via
the local probability model, \emph{the LqHV probability model }\cite{6, 21},
incorporating the LHV model only as a particular case. An arbitrary
correlation scenario admits a LqHV model if and only if it is nonsignaling.

The LqHV modelling framework allowed us to construct the nonsignaling
analogs (\ref{18}) of Bell inequalities and to specify parameters (\ref{23}%
), (\ref{24}) quantifying Bell's nonlocality, partial and full, in a general
nonsignaling case. For quantum correlation scenarios on an arbitrary $N$%
-partite quantum state, we evaluate these nonlocality parameters
analytically (\ref{27}) in terms of dilation characteristics of an $N$%
-partite state. For an $N$-qudit state, we also evaluate these parameters
numerically (\ref{32}), (\ref{34}) in $d,$ $S$ and $N.$

We stress that, in a LqHV model (\ref{12}), locality is introduced quite
similarly as in a LHV model (\ref{5}) and the only difference between these
two local probability models is nonpositivity of a normalized real-valued
distribution $\mu _{\mathcal{E}_{S,\Lambda }}^{lqhv}$ in the general LqHV
representation (\ref{12}). Therefore, a LHV model constitutes a particular
case of a LqHV model whenever a distribution $\mu _{\mathcal{E}_{S,\Lambda
}}^{lqhv}$ is positive.

From the mathematical modelling point of view, a LqHV model reproduces
correctly all scenario joint probability distributions, so that it is not
important that a simulation distribution $\mu _{\mathcal{E}_{S,\Lambda
}}^{lqhv}$ may have negative values -- for observed events, there are no
negative probabilities.

However, from the conceptual point of view, it is important to understand a
reason of nonvalidity of an LHV model in a general nonsignaling case,
specifically, in a quantum case.

Bell argued \cite{2, 3} that violation of the LHV\ statistical constraints
under space-like separated quantum measurements points to violation of the
EPR locality \cite{30} under these measurements. However, as we analyzed
mathematically in section 3 of \cite{5}, for an arbitrary correlation
scenario, Bell's nonlocality does not imply violation of the EPR locality.
Moreover, under space-like separated quantum measurements, the EPR locality
is not violated.

Furthermore, in view of the existence of the local probability model, \emph{%
the LqHV model }\cite{6, 21}, correctly reproducing the Hilbert space
description of every quantum correlation scenario, we argued in \cite{7}
that quantum violation of Bell inequalities can be explained not by
violation of the EPR locality but by \emph{nonclassicality} leading to
violation of "classical realism" embedded into a HV model via a \emph{%
probability} distribution. In a LqHV model, the locality is preserved but
"classical realism" is replaced by "quantum realism" modeled by a
real-valued distribution.

The results of the present paper on the rigorous mathematical description of
Bell's nonlocality in a general nonsignaling case via \emph{the local
probability model} confirm our opinion gradually formed in \cite{4, 5, 6, 7,
8} -- \emph{violation of Bell inequalities in a quantum case is not due to
violation of the EPR\ locality conjectured by Bell \cite{2, 3} but due to
the improper HV modelling of "quantum realism".}

In conclusion, we stress that the Kolmogorov probability axioms \cite{31}
refer to the probabilistic description of a single measurement and are true
for a measurement of any type, classical or quantum. But a correlation
scenario is described by a family of joint measurements and, though, for
each joint measurement, the Kolmogorov probability axioms are fulfilled, the
probabilistic description of the whole correlation scenario does not need to
be described in terms of a single probability space inherent to the
Kolmogorov probability model. In quantum theory, the Kolmogorov probability
model is referred to as the HV model.

However, as we proved in \cite{6, 21}, the probabilistic description of
every nonsignaling correlation scenario, in particular, every quantum
correlation scenario, does admit the LqHV probability model. By taking off
the specification on locality in the LqHV model and generalizing it for the
probabilistic description of an arbitrary measurement situation, we
introduced in \cite{21} the notion of \emph{the} \emph{qHV (quasi hidden
variable) model }--\emph{\ }equivalently, the quasi-classical probability
model -- \emph{\ }incorporating the Kolmogorov probability model only as a
particular case. We proved \cite{22} that, in its context-invariant form,
the qHV model reproduces the probabilistic description of all joint von
Neumann measurements on an arbitrary Hilbert space.\bigskip

\subparagraph{\noindent \textbf{Acknowledgements.}}

The valuable discussions with Professor A. Khrennikov are very much
appreciated.

\end{document}